\documentclass[aps,prb,reprint,superscriptaddress,showpacs,showkeys,preprintnumbers,
nofootinbib,
amsmath,amssymb,
aps,
prb,
]{revtex4-2}

\usepackage{graphicx}
\usepackage{dcolumn}
\usepackage{bm}
\usepackage{mathrsfs}
\usepackage[T1]{fontenc}
\usepackage{scalerel}

\usepackage{siunitx}
\usepackage{physics}
\AtBeginDocument{\RenewCommandCopy\qty\SI}
\DeclareSIUnit\rydberg{\text{Ry}}
\DeclareSIUnit{\au}{a.u.}
\usepackage[normalem]{ulem} 
\usepackage[dvipsnames]{xcolor}

\begin{document}
	
\author{Lifen~Xiang }
\affiliation{Beijing National Laboratory for Condensed Matter Physics, Institute of Physics, Chinese Academy of Sciences, Beijing 100190, China}

\author{Siyi~Lei}
\affiliation{Beijing National Laboratory for Condensed Matter Physics, Institute of Physics, Chinese Academy of Sciences, Beijing 100190, China}
\author{Xiaolin~Ren}
\affiliation{Beijing National Laboratory for Condensed Matter Physics, Institute of Physics, Chinese Academy of Sciences, Beijing 100190, China}
\author{Ziao~Han}
\affiliation{Beijing National Laboratory for Condensed Matter Physics, Institute of Physics, Chinese Academy of Sciences, Beijing 100190, China}
\affiliation{University of Chinese Academy of Sciences, Beijing 100049, China}
\author{Zijian~Xu}
\affiliation{Shanghai Synchrotron Radiation Facility, Shanghai Advanced Research Institute, Chinese Academy of Sciences, Shanghai 201204, China}
\author{X.J.~Zhou}
\email{XJZhou@iphy.ac.cn}
\affiliation{Beijing National Laboratory for Condensed Matter Physics, Institute of Physics, Chinese Academy of Sciences, Beijing 100190, China}
\affiliation{Beijing Academy of Quantum Information Sciences, Beijing 100190, China}
\affiliation{Songshan Lake Materials Laboratory, Dongguan 523808, China}
\author{Zhihai~Zhu}
\email{zzh@iphy.ac.cn}
\affiliation{Beijing National Laboratory for Condensed Matter Physics, Institute of Physics, Chinese Academy of Sciences, Beijing 100190, China}
\affiliation{Songshan Lake Materials Laboratory, Dongguan 523808, China}

\title{Stabilizing and Tuning Superconductivity in La$_3$Ni$_2$O$_{7-\delta}$ Films: Oxygen Recycling Protocol Reveals Hole-Doping Analogue }
	
\begin{abstract}
	The recent achievement of superconductivity in La$_3$Ni$_2$O$_{7-
		\delta}$ with transition temperatures exceeding 40 K in thin films under compressive strain and 80 K in bulk crystals under high pressure opens new avenues for research on high-temperature superconductivity. The realization of superconductivity in thin films requires delicate control of growth conditions, which presents significant challenges in the synthesis process. Furthermore, the stability of superconducting La$_3$Ni$_2$O$_{7-\delta}$ films is compromised by oxygen loss, which complicates their characterization. We introduce an effective recycling protocol that involves oxygen removal in a precursor phase followed by ozone-assisted annealing, which restores superconducting properties. By tuning the oxygen content, we construct an electronic phase diagram that highlights oxygen addition as a potential analogue to hole doping via La substitution with Sr, providing insights into the doping mechanism and guiding future material optimization. 
\end{abstract}

\maketitle

\section{Introduction}
	The discovery of superconductivity in La$_3$Ni$_2$O$_{7-
		\delta}$ at 14 GPa with a $T_c$ of around 80 K has recently garnered considerable attention~\cite{sunSignaturesSuperconductivity802023}, leading to the exploration of compressive strain as an alternative to pressure by growing thin films on mismatched substrates. Superconductivity has been achieved in thin films of La$_3$Ni$_2$O$_{7-
		\delta}$, as well as in Pr- and Sr-doped variants, with onset transition temperatures ranging from 42 to 48 K~\cite{koSignaturesAmbientPressure2025,liuSuperconductivityNormalstateTransport2025,zhouAmbientpressureSuperconductivityOnset2025,haoSuperconductivityPhaseDiagram2025}. Although these films have a much lower $T_c$ than the maximum $T_c$ achieved by applying high pressure to bulk crystals~\cite{sunSignaturesSuperconductivity802023,Hou_2023,zhangHightemperatureSuperconductivityZero2024,wangPressureInducedSuperconductivityPolycrystalline2024,shiPrerequisiteSuperconductivitySDW2025,liAmbientPressureGrowth2025,wangBulkHightemperatureSuperconductivity2024}, so far, the realization of superconductivity in thin films provides excellent opportunities to investigate the mechanism of superconductivity in La$_3$Ni$_2$O$_{7-
		\delta}$, as these films can exhibit superconductivity without high pressure and are more compatible with various experimental techniques~\cite{wangElectronicStructureCompressively2025,10.1093/nsr/nwaf205}.
        
		However, synthesizing superconducting La$_3$Ni$_2$O$_{7-\delta}$ films requires a very narrow growth window, an extremely thin thickness, and an in situ or post-annealing process with ozone. This creates significant challenges for producing superconducting samples suitable for advanced measurements. Additionally, these superconducting films are sensitive to air and tend to lose oxygen, becoming nonsuperconducting after exposure to air~\cite{zhouAmbientpressureSuperconductivityOnset2025}. These issues emphasize the importance of evaluating reproducibility and stability, as well as developing effective methods for recycling these La$_3$Ni$_2$O$_{7-\delta}$ films when they degrade. Furthermore, recycling the same film multiple times can help identify the main factors influencing superconductivity, providing insights into superconductivity from a sample synthesis perspective. 
        
		In this study, we report the achievement of superconductivity in La$_3$Ni$_2$O$_{7-\delta}$ films on SrLaAlO$_4$(001) (SLAO) via pulsed laser deposition (PLD), subsequently followed by ozone annealing. A practical methodology for the recycling of degraded superconducting films employing a two-step annealing process, namely, oxygen removal and ozone annealing, is presented. This process allows a single film to transition between insulating and superconducting states reversibly. Furthermore, by subjecting a single film to ozone annealing, an electronic phase diagram is proposed in which oxygen addition may exert an effect analogous to hole doping by substituting La for Sr~\cite{haoSuperconductivityPhaseDiagram2025}. Finally, X-ray absorption spectroscopy (XAS) conducted on the as-grown film on a SLAO substrate indicates the presence of holes within the Ni $d_{z^2}$-derived bonding band, similar to what is observed in a thick film grown on the LaAlO$_3$ (LAO) substrate~\cite{renResolvingElectronicGround2025}. 

\section{Results}
\begin{figure}[t]
	\centering
	\includegraphics[width=.50\textwidth]{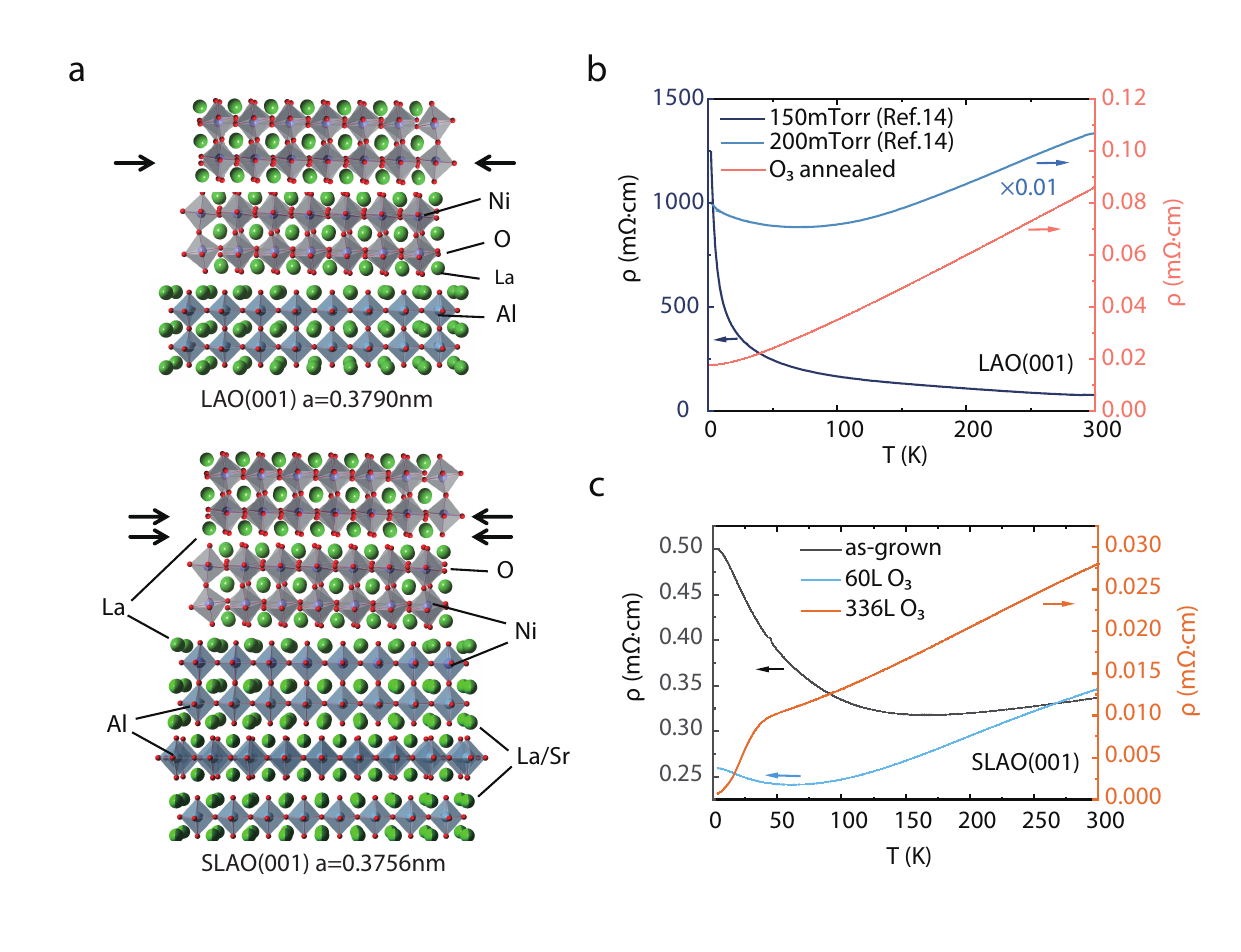}
	\caption{{\bf Schematic structural and electrical characterization of La$_3$Ni$_2$O$_{7-\delta}$ films on LaAlO$_3$(001)(LAO) and SrLaAlO$_4$(001)(SLAO) substrates.} {\bf a}, the schematic structure of La$_3$Ni$_2$O$_{7-\delta}$ films grown on LAO and SLAO substrates. {\bf b}, $\rho$(T) curves for three typical La$_3$Ni$_2$O$_{7-\delta}$ films with a thickness of ~36 nm on LAO substrates with varying oxygen pressure during the growth, where the $\rho$(T) curve of the film grown under 150 mTorr and annealed in ozone is shown in red. {\bf c}, $\rho$(T) curves for three typical 5.5 nm-thick La$_3$Ni$_2$O$_{7-\delta}$ films grown on SLAO substrates with different ozone annealing treatments, where the curve in orange shows that superconducting transition occurs at $\sim$ 40 K. }
	\label{fig:fig1}
\end{figure}

	Compressive strain and oxygen stoichiometry are key factors in controlling the electronic properties of La$_3$Ni$_2$O$_{7-\delta}$ films~\cite{renResolvingElectronicGround2025}. Figure 1a shows single-crystalline La$_3$Ni$_2$O$_{7-\delta}$ thin films grown on two substrates, LAO and SLAO, with SLAO inducing $\sim$ 0.9\% compressive strain compared to LAO based on their lattice constants. Figure 1b displays resistivity measurements of three typical La$_3$Ni$_2$O$_{7-\delta}$ films grown on LAO with different oxygen stoichiometries. As the oxygen content increases, the films transition from insulators to metals, although they do not exhibit superconductivity. However, using the SLAO substrate along with an oxygen annealing process, as shown in Fig. 1c, enables the films to evolve from insulators to metals and ultimately to superconductors. The presence of Sr due to the SLAO substrates may also contribute to the development of superconductivity at the interface between the film and the substrate~\cite{zhouAmbientpressureSuperconductivityOnset2025}. However, since superconductivity in pressurized bulk crystals does not involve Sr, the presence of Sr from substrates, if relevant, may only cause variations in hole concentration by substituting La. This is consistent with the recent study on Sr-doped La$_3$Ni$_2$O$_{7-\delta}$ films~\cite{haoSuperconductivityPhaseDiagram2025}.
	
	To further confirm the transition in resistivity measurements shown in Fig. 1c, exhibiting superconducting behavior, we investigate the transport properties with a magnetic field (up to 9.0 Tesla) applied perpendicular to the ab-plane of two typical films using a Quantum Design Physical Property Measurement System (PPMS) with a standard four-probe configuration. Figures 2a and 2b show that the field dependence of resistivity measurements for both films follows a similar pattern: in the normal state, the films exhibit minimal variation; however, below the superconducting transition temperature, which occurs at $T_c$ $\approx$ 42K, the conductance is significantly suppressed as the magnetic field increases. We then estimate the upper critical field $H_{c,\perp}$ for the magnetic field perpendicular to the ab-plane of the film based on $T_c$ at 90\% and 50\% of the resistance in the normal state near the onset of $T_c$, using the linearized Ginzburg–Landau form written as
	
	\begin{equation}
		H_{c,\perp} = \frac{\phi_0}{2\pi\xi^2_{ab}}(1-T/T_c) 
	\end{equation}
	
	where $\phi_0$ is the flux quantum, $\xi_{ab}$ is the zero-temperature Ginzburg–Landau coherence length. The extracted data and fitting are presented in Figs.2c and 2d, which correspond to the data displayed in Figs. 2a and 2b, respectively. Both films exhibit comparable $H_{c, 50\%}$$\approx$ 16.9 $\pm$ 0.6 (19.3 $\pm$ 0.7)T, while the extracted value for $H_{c, 90\%}$ in Fig. 2c is 103.4 $\pm$ 2.8T, and in Fig. 2d it is 58.4 $\pm$ 1.5T. This difference may stem from the sample qualities, as noted, with a second transition occurring in the formal case, highlighted by the arrow in Fig. 2c, as well as the different thicknesses. In general, our results align with the values reported in the references~\cite{koSignaturesAmbientPressure2025}. We found that $\xi_{ab}$ is approximately 2.67nm and 2.37nm for the film in Figs. 2a and 2b, which are similar to the reported values for the film~\cite{koSignaturesAmbientPressure2025}  and bulk~\cite{zhangHightemperatureSuperconductivityZero2024}. {\color{black}This short coherence length resembles observations made in cuprates, suggesting that short-range Coulomb repulsion is likely an important factor in understanding the electronic structures and superconductivity of La$_3$Ni$_2$O$_{7-\delta}$. This is similar to the findings in cuprates, where strong local minima in the Coulomb interaction are observed within the range of 0.5 to 1 nm~\cite{derricheAtomicBondPolarization2025}. \begin{figure}[t]
		\centering
		\includegraphics[width=0.5\textwidth]{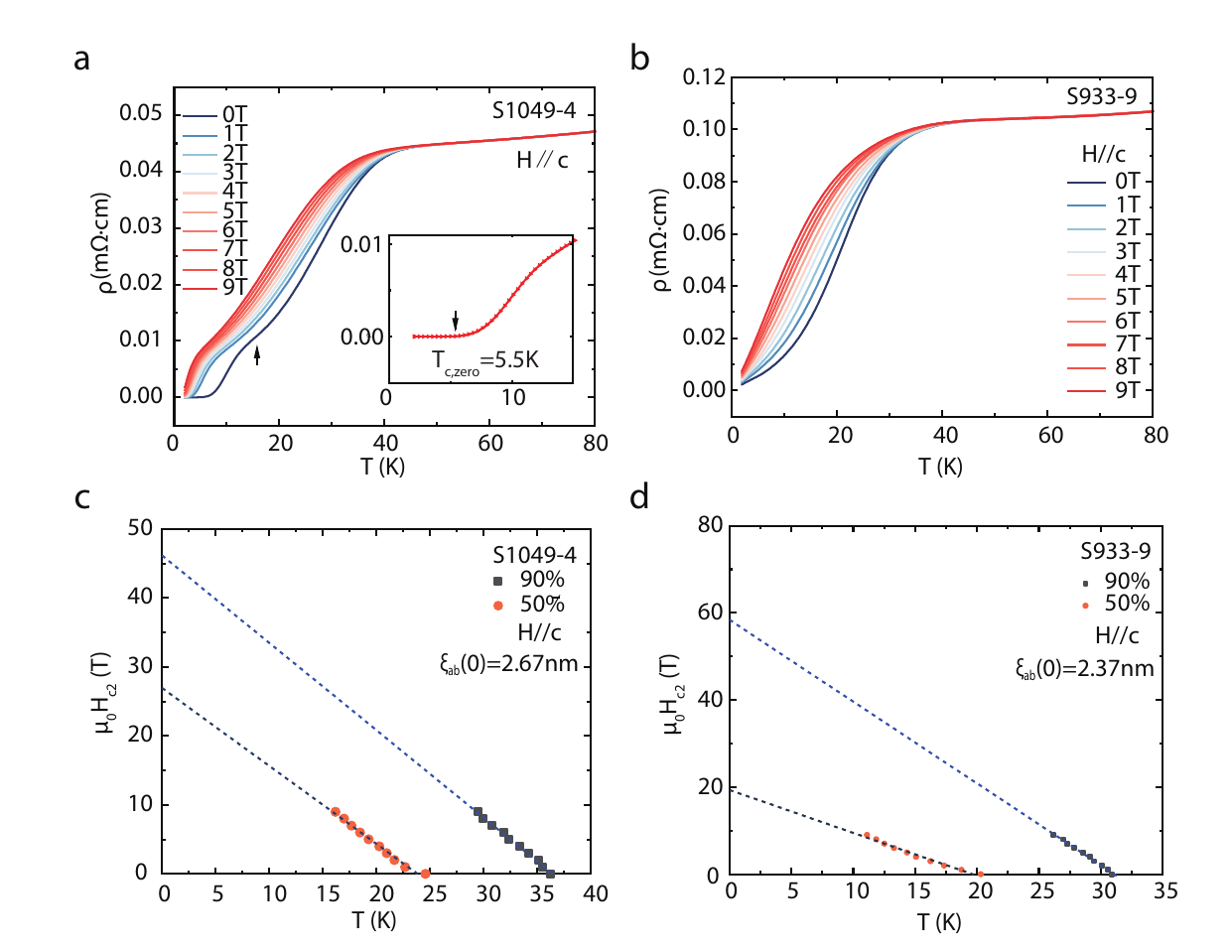}
		\caption{ {\bf Transport properties of La$_3$Ni$_2$O$_{7-\delta}$ thin films on SLAO substrates under a magnetic field.} {\bf a, b,} $\rho$(T) curves under various magnetic fields applied perpendicular to the $ab$ plane of films with different thicknesses, 8.5 nm and 3.5 nm, respectively; {\color{black}{the arrow in Fig. 2a indicates that a second transition emerges at $\sim$18 K, and the inset shows a zoomed-in view of resistivity measurements demonstrating that zero resistance occurs below 5.5 K.}}  {\bf c,d,} Solid circles and squares represent the upper critical fields ($H_{c,\perp}^{50\%}$ and $H_{c,\perp}^{90\%}$) extracted by the $T_{c,50\%}$ and $T_{c,90\%}$.   Dotted lines are Ginzburg-Landau fits.}
		\label{fig:fig2}
	\end{figure} 
	\begin{figure*}[thb]
		\centering
		\includegraphics[width=.9\textwidth]{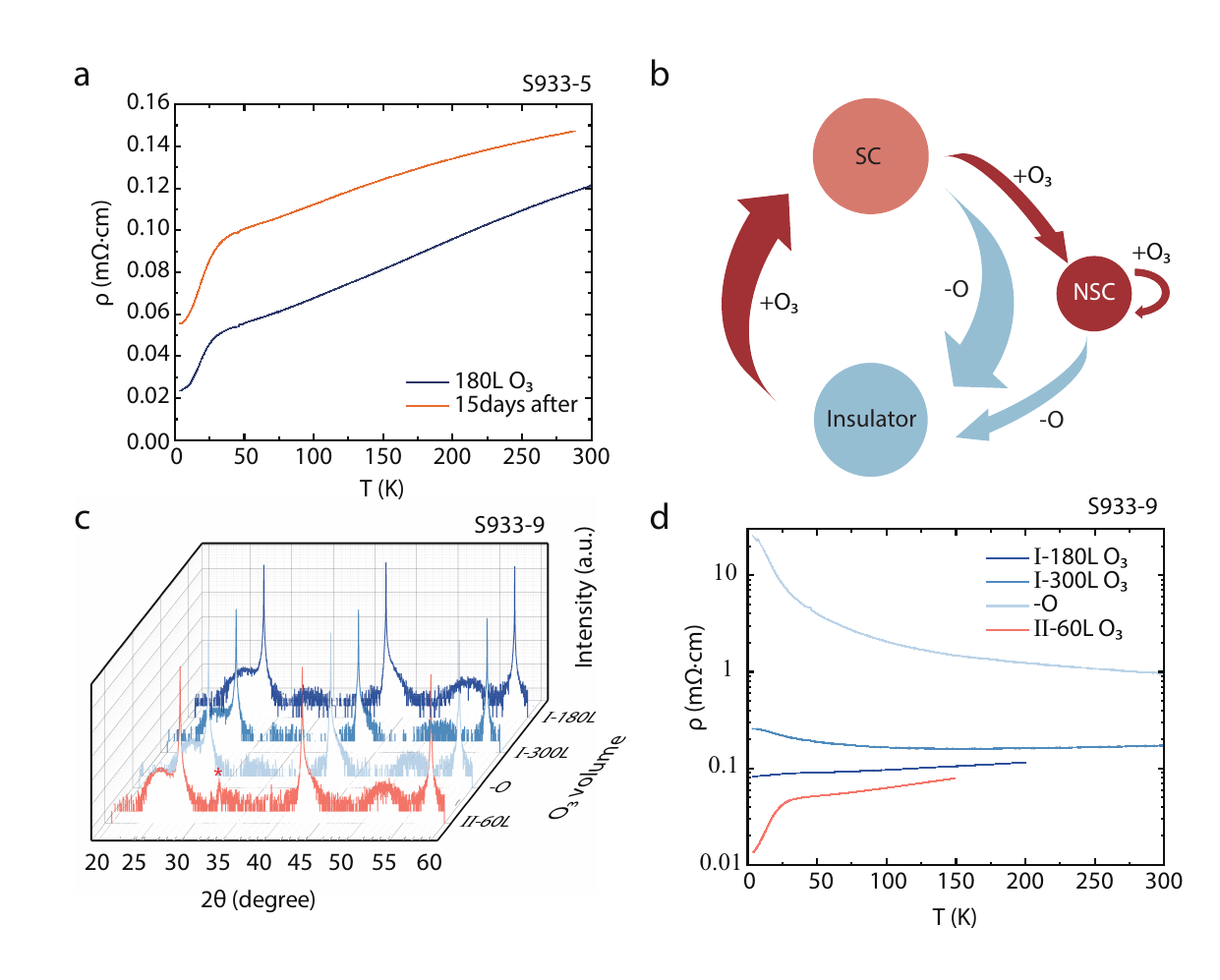}
		\caption{{\bf Recycle of degraded superconducting La$_3$Ni$_2$O$_{7-\delta}$ thin films.} {\bf a,} $\rho$(T) curves for a typical superconducting La$_3$Ni$_2$O$_{7-\delta}$ film with a thickness of 3.5 nm, and the same film stored in an argon-filled glove box for 15 days. {\bf b,} Illustration of the two-step process for recycling the degraded superconducting La$_3$Ni$_2$O$_{7-\delta}$ films; Reannealing the degraded superconducting films under ozone tends to destroy the superconductivity. However, the two-step process, i.e., removing oxygen followed by an annealing process under ozone, can effectively restore the superconducting signature in the film. {\bf c, d,}The out-of-plane XRD pattern and $\rho$(T) curves for a typical 3.5 nm-thick La$_3$Ni$_2$O$_{7-\delta}$ thin film under recycling.{\color{black}{The mixed-gas volume is estimated by multiplying the flow rate by the total annealing time. It flows continuously into a chamber of about 0.5 L for annealing, as ozone lifetime above 100$^\circ$C is short.}}}
		\label{fig:fig3}
	\end{figure*}

	We demonstrate an effective method for recycling degraded superconducting La$_3$Ni$_2$O$_{7-\delta}$ films. As shown in Fig. 3a, the film that displays a superconducting signature shows a generally higher resistance after 15 days in the argon-filled glovebox, although the superconducting transition remains clear. One might naively think that the easiest way to recycle these degraded superconducting films is to reanneal them in ozone, as was initially done for the as-grown films. However, this method is shown to harm superconductivity, ultimately turning the film into an insulator or metal and potentially damaging its structure. Instead, we first anneal the degraded films in air to remove oxygen, then anneal them in ozone to reintroduce oxygen. We have found that this two-step process, summarized in Fig. 3b, is highly effective at restoring superconductivity in degraded films, allowing the same film to be recycled multiple times. During recycling, we analyze the crystal structure of a typical film at each stage by using its X-ray diffraction (XRD) pattern along with transport measurements. For example, we show the evolution of the out-of-plane XRD pattern and the corresponding resistivity measurements on a typical film in Figs. 3c and 3d, respectively. In Fig. 3c, it is clear that the diffraction peaks characteristic of the La$_3$Ni$_2$O$_{7-\delta}$ phase weaken as the film degrades. After applying the two-step process, the diffraction peaks regain their original intensity, and the film becomes a superconductor, as indicated by the resistivity measurements in Fig. 3d. {\color{black}Taken together, the main causes of film degradation may be oxygen loss and structural deterioration. The two-step process can effectively reintroduce oxygen and prevent over-oxidation of the degraded films, whose oxygen stoichiometry stays near the level required for superconductivity. Additionally, removing oxygen could improve and strengthen the crystalline structure of the degraded films.} 
    
	\begin{figure*}[thb]
		\centering
		\includegraphics[width=.95\textwidth]{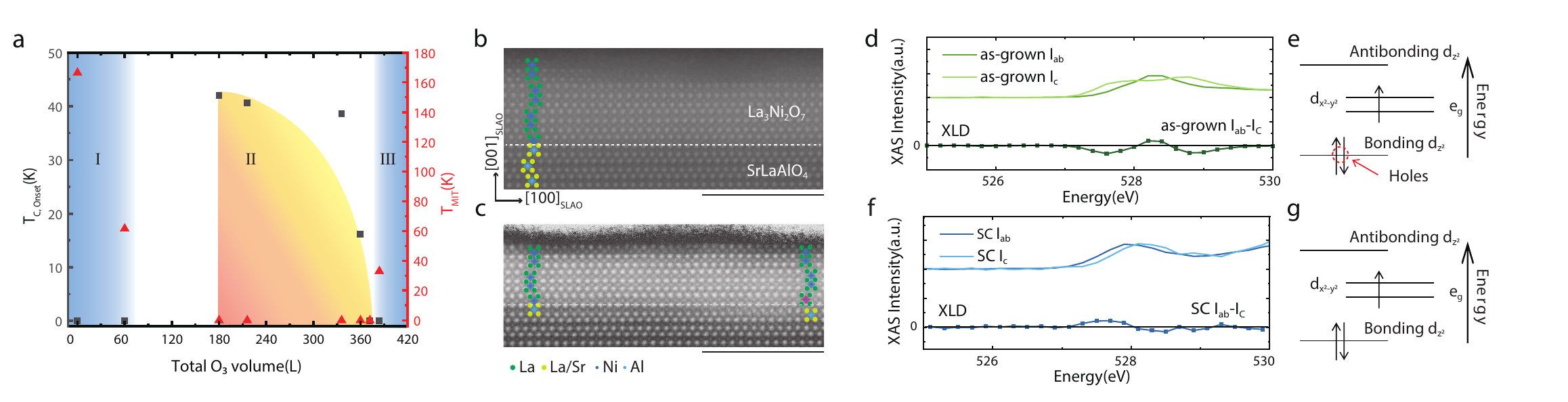}
		\caption{{\bf Phase diagram, STEM image, and XAS of La$_3$Ni$_2$O$_{7-\delta}$ thin film.} {\bf a,}The solid squares and triangles represent the $T_{c,onset}$, and the $T_{MIT}$, extracted from $\rho$(T) curves after each annealing with different $O_3$ flow rates. We caution that the horizontal coordinate indicates the total volume of $O_3$ used during the annealing process, which might not precisely represent the actual oxygen content in the film. Instead, it indicates that oxygen gradually enters the film, meaning $\delta$ decreases as it is annealed step-by-step in ozone. {\bf b, c,} ADF-STEM images of 3.5 nm cycled La$_3$Ni$_2$O$_{7-\delta}$ thin film on SLAO at various regions along the [010] axis of the substrate. The La$_3$Ni$_2$O$_{7-\delta}$ thin film-substrate interface is marked by the white dashed line. The reduced intensity near the surface may result from damage during sample preparation for STEM. The positions of La, Ni, La/Sr, and Al ions are indicated by different colors. At the film-substrate interface, the bilayer (LaSr)$_3$Ni$_2$O$_{7-\delta}$ and monolayer La$_2$NiO$_4$ structures are clearly visible. The octahedra in La$_2$NiO$_4$ are highlighted in red. Scale bars are 5 nm in (b-c). {\bf d,} XAS measurements on a typical film with $\sim$7 nm thickness grown on SLAO, suggests the presence of holes in Ni $d_{z^2}$ derived bonding band, as illustated in {\bf e}. {\bf f,} XAS measurements on a superconducting film with $\sim$8.5 nm thickness, reveal only two components corresponding to $d_{x^2-y^2}$ and $d_{z^2}$ anti-bonding states, indicating the absence of holes in Ni $d_{z^2}$ derived bonding band, as illustrated in {\bf g}. }
		\label{fig:fig4}
	\end{figure*}
	
	We propose an electronic phase diagram in Fig. 4a for La$_3$Ni$_2$O$_{7-\delta}$ by adding oxygen to the films. First, we remove oxygen from a film by annealing it in air, creating a state referred to as the precursor phase, which is an insulating state. Starting from this insulating state, we gradually added oxygen to the film by annealing it in an ozone environment. At the same time, we examine the crystal structure of the film at each step to ensure that it remains structurally intact and preserves the La$_3$Ni$_2$O$_{7-\delta}$ phase, with $\delta$ being the primary variable during annealing. In this way, the introduction of oxygen, reducing $\delta$—can mimic hole doping through the substitution of La with Sr. As shown in Fig. 4a, three typical regimes are identified from the transport measurements. In the first regime, the initial film is insulating and becomes metallic upon oxidation. However, at low temperatures, the resistivity versus temperature shows an upturn, which we call the metal-to-insulator transition. We define the corresponding transition temperature as $T_{MIT}$, as shown in the phase diagram. At a critical doping level achieved by ozone annealing, the film exhibits superconducting behavior, with $T_{c,onset}$ decreasing with increasing oxygen content. Once a new critical oxygen level is reached, the film loses its superconductivity and returns to a metallic state; however, it exhibits an upturn at low temperatures, similar to the first regime. We caution that we cannot measure oxygen concentration precisely. In the phase diagram, equal intervals are based on the cumulative volume of O$_3$ used in the annealing procedure, with each stage fixed at 2 hours. Adding oxygen to the film can become increasingly difficult, and the actual amount of oxygen added decreases as $\delta$ approaches its minimal value. {\color{black}{Besides, $T_{c,zero}$ appears much lower and harder to reach in La$_3$Ni$_2$O$_{7-\delta}$ films than in Pr- and Sr-doped films~\cite{zhouAmbientpressureSuperconductivityOnset2025,liuSuperconductivityNormalstateTransport2025,haoSuperconductivityPhaseDiagram2025}, probably owing to sample details such as impurity phase, sample granularity, nonuniformity, or interface effects. In contrast, $T_{c, onset}$ for a film of La$_3$Ni$_2$O$_{7-\delta}$ subjected to similar O$_3$ annealing often remains comparable regardless of whether T$_{c,zero}$ is achieved.}}  We observe a phase diagram similar to that reported recently in Sr-doped films~\cite{haoSuperconductivityPhaseDiagram2025}. In both cases, the superconducting region has an asymmetric shape, sharply contrasting with the more symmetric superconducting dome seen in hope-doped cuprate superconductors~\cite{keimerQuantumMatterHightemperature2015b}.
	
	However, the details of the La$_3$Ni$_2$O$_{7-\delta}$ samples are much more complex, and changing $\delta$ by annealing is probably not the only factor that affects $T_c$ and its onset. In particular, as shown in Fig. 3c, we detected unknown phases with a moderately sharp diffraction peak at 32.9$^{\circ}$ that appear during annealing, and the film containing these phases still exhibits superconductivity. We can exclude the possibility that this unknown phase originates from the substrate because a blank substrate subjected to a similar two-step annealing process does not produce a peak at 32.9$^{\circ}$ (see the Supplementary Materials for details). This suggests that the unknown phase is likely related to the film itself or the interface between the film and the substrate. The Ruddlesden-Popper (RP) series (La$_{n+1}$Ni$_n$O$_{3n+1}$) may be a candidate for this unknown phase, as their (110) Bragg peaks match 2$\theta$ $\approx$ 32.9$^{\circ}$. The unknown impurity phases could explain a very broad transition, resulting in a low $T_{c, zero}$. Additionally, considering that superconductivity occurs only in the ultrathin limit, interface reconstruction may play a role. In Fig. 4b, we show the scanning tunneling microscopy image of the dominant phase of a superconducting film with the (2222)-type interface formed by the top layers of the substrate and the first epitaxial layer of the film. In Fig. 4c, we show a typical region that consists of the same interface structure shown in Fig. 4b, along with an interface structure that forms a single-layer phase of type (214), followed by an epitaxially grown bilayer phase of La$_3$Ni$_2$O$_{7-\delta}$. Our observations on the interface structures align well with earlier studies~\cite{koSignaturesAmbientPressure2025}. In our films, we did not observe the (1313)-type polymorph, which is present in bulk crystals~\cite{puphalUnconventionalCrystalStructure2024,chenPolymorphismRuddlesdenPopper2024,wangStructureResponsibleSuperconducting2024}.   
	
	Figures 4d and 4f show photon-polarized XAS measurements on as-grown and superconducting films of La$_3$Ni$_2$O$_{7-\delta}$ deposited on SLAO substrate. {\color{black}The X-ray linear dichroism (XLD) in Fig. 4d reveals a similar pattern as in the LAO substrated film: a peak with in-plane polarization along with two dips with out-of-plane polarization~\cite{renResolvingElectronicGround2025}. It suggests that the choice of substrates and film thickness does not significantly alter the electronic ground state of Ni ions in as-grown films. In the as-grown films, regardless of the substrates or thickness, the XLD measurements indicate that there may still be unoccupied states in Ni $d_z^2$-derived bonding states, as illustrated in Fig. 4e. In superconducting films grown on SLAO substrates, the XLD analysis shows that the spectral weight from $d_z^2$ bonding states is nearly absent, as shown in Fig. 4f. The corresponding orbital configurations of Ni ions are illustrated in Fig. 4g, clearly contrasting with the observations in as-grown and non-superconducting films.} 
	
	\section{Discussion}
	Considerable progress has been made in understanding the origin of superconductivity in this compound  La$_3$Ni$_2$O$_{7-\delta}$~\cite{yangOrbitaldependentElectronCorrelation2024,au-yeungUniversalElectronicStructure2025,luoBilayerTwoOrbitalModel2023,zhangElectronicStructureDimer2023,liuWavePairingDestructive2023,lechermannElectronicCorrelationsSuperconducting2023,yangPossibleWaveSuperconductivity2023,guEffectiveModelPairing2025,shenShenEffectiveBiLayerModel2023,sakakibaraPossibleHighSuperconductivity2024,shilenkoCorrelatedElectronicStructure2023,luInterlayerCouplingDrivenHighTemperatureSuperconductivity2024,quBilayerModelMagnetically2024,yashimaMicroscopicEvidenceSpin2025,fengUnconventionalSuperconductingPairing2025,zhanImpactNonlocalCoulomb2025,ohHighSpinLow2025,yiUnifyingStraindrivenPressuredriven2025,kaneko$t$$J$ModelStrongly2025}. In particular, films that can superconduct under ambient pressure and are compatible with various spectroscopic techniques have led to significant progress, allowing the direct measurement of the electronic structures and details of the excitation spectrum of superconducting films~\cite{shenNodelessSuperconductingGap2025,wangElectronicStructureCompressively2025,10.1093/nsr/nwaf205,fanSuperconductingGapStructure2025,sunObservationSuperconductivityinducedLeadingedge2025}. The electronic ground state of Ni ions in La$_3$Ni$_2$O$_{7-\delta}$, which is fundamental to many theoretical models, continues to be a topic of debate. Our XAS measurements on an as-grown film of La$_3$Ni$_2$O$_{7-\delta}$ with a thickness of $\sim$7 nm on the SLAO substrate suggest the presence of holes in the Ni $d_{z^2}$-derived bonding bands, which is consistent with the observations in the 36-nm thick films grown on LAO~\cite{renResolvingElectronicGround2025}. {\color{black}{In contrast, in a superconducting film grown on SLAO with a thickness of $\sim$ 8.5 nm, the XAS measurements indicate that the spectral weight from the Ni $d_{z^2}$-derived bonding bands is significantly suppressed~\cite{wangElectronicStructureCompressively2025,sunObservationSuperconductivityinducedLeadingedge2025}.}} However, this interpretation assumes nearly stoichiometric conditions, namely that $\delta$ is close to zero in La$_3$Ni$_2$O$_{7-\delta}$. We cannot rule out that oxygen vacancies induce distinct local crystal fields, leading to two distinct Ni 3d$_{z^2-3r^2}$ energies. Moreover, naively, adding extra oxygen through ozone annealing is expected to introduce more holes into the sample, ultimately promoting superconductivity. However, our recycling protocol indicates that there is no continuous increase in superconducting transition temperatures, unlike cuprate superconductors, where the addition of holes gradually increases $T_c$ to its maximum, except that near 1/8 doping region, superconductivity is suppressed by charge or stripe order in the electronic phase diagram~\cite{keimerQuantumMatterHightemperature2015b}. This suggests a potentially more complex role for oxygen in transforming as-grown or recycled films with reduced oxygen into superconductors. There are likely significant oxygen deficiencies in the as-grown film or the oxygen-reduced film~\cite{dongInterstitialOxygenOrder2025,haoSuperconductivityPhaseDiagram2025}. Starting from the oxygen-depleted phase~\cite{foyevtsovaChargeDistributionMagnetism2025}, the added oxygen can initially fill the {\color{black}oxygen vacancies, which are likely located at the inner apical positions since the apical oxygen here governs the bonding and antibonding states derived from the $d_z^2$ orbitals, aligning with our XAS measurements that show the $d_z^2$ derived bonding states change significantly between as-grown and superconducting films.} Once superconductivity is realized, subsequent oxygen supplementation may predominantly modify the hole concentration, thereby progressively decreasing $T_c$, analogous to hole doping through substituting La by Sr~\cite{haoSuperconductivityPhaseDiagram2025}. {\color{black}{It is acknowledged that, in bulk crystals, oxygen may occupy interstitial sites, thereby introducing oxygen periodicity and competing with superconductivity~\cite{dongInterstitialOxygenOrder2025}, whereas in thin films of Sr-doped bilayer nickelates, oxygen vacancies predominantly reside within the NiO$_2$ planes~\cite{haoSuperconductivityPhaseDiagram2025}. The role of added oxygen， in addition to its influence on charge-carrier concentration in La$_3$Ni$_2$O$_{7-\delta}$ films during annealing， remains incompletely understood and warrants further investigation.}}   
	
	\section{Conclusion}

	This study presents an effective method for recycling degraded superconducting La$_3$Ni$_2$O$_{7-\delta}$ films, using a two-step process: oxygen removal in ambient air followed by annealing in ozone. This approach allows for precise control over a single film, enabling the reversible transition between superconducting and non-superconducting states, which could help identify the key features responsible for superconductivity. Additionally, by gradually adding oxygen to the film, a phase diagram is proposed that may simulate hole doping of the film through La substitution with Sr.

	\noindent\textbf{Methods}\\
	\noindent The thin films of La$_3$Ni$_2$O$_{7-\delta}$ were grown on SrLaAlO$_4$ substrates using pulsed laser deposition (PLD), described in ref.~\cite{renResolvingElectronicGround2025}. The superconducting transition temperature was examined by transport measurements using a Quantum Design Physical Property Measurement System (PPMS) with a standard four-probe configuration. Samples for cross-sectional scanning transmission electron microscopy (STEM) were prepared using focused ion beam (FIB, Helios 600i) techniques. The high-angle annular dark field (HAADF) images were acquired on the ARM-200F (JEOL, Japan) operated at 200 kV with a CEOS Cs corrector (CEOS GmbH, Germany). XAS experiments were carried out at T = 17 K at Beamline BL08U1A of Shanghai Synchrotron Radiation Facility (SSRF). The absorption spectra were collected using the total electron-yield (TEY) mode with linear vertical ($\sigma$) and horizontal ($\pi$) light polarizations.

\noindent\textbf{Acknowledgments}\\
	We thank George Sawatzky and Xianxin Wu for their insightful discussions and thorough comments on the manuscript. We also acknowledge Yu Zhang for experiment assistance on XAS measurements. This work was supported in part by the National Key Research and Development Program of China (Grant No. 2022YFA1403900 and 2021YFA1401800), the National Natural Science Foundation of China (Grant No. 12494593), the Strategic Priority Research Program (B) of the Chinese Academy of Sciences (Grant No. XDB25000000), CAS Superconducting Research Project (Grant No. SCZX-0101) and the Synergetic Extreme Condition User Facility (SECUF). \\
	
\noindent\textbf{Author contributions}\\
	X.J.Z. and Z.H.Z. conceived the project. L.X., S.L., and X.R. grew and characterized the films.
	X.L., Z.H., Z.X., and Z.H.Z. carried out the XAS experiments at 08L, SSRF. L.X. and Z.H.Z. wrote the manuscript with input from other authors.\\

\noindent\textbf{Data availability} \\
Data supporting the findings of this study are available from the corresponding authors on a reasonable request.\\
		
\bibliography{Recycling}

\end{document}